\documentstyle[12pt,moriond,psfig]{article}
\begin{document}
\heading{Dwarfmology}

\author{Simon P. Driver and Alberto Fern\'andez-Soto} {School of Physics, University of New South Wales,}{Sydney, NSW 2052, AUSTRALIA.}

\begin{moriondabstract}
We address two important cosmological questions relating to dwarf galaxies: 
(1) What is their contribution to the field galaxy baryon budget; and (2) 
what is their contribution to the faint blue galaxy problem. 
Both of these are addressed empirically from a complete photometric redshift 
catalogue\cite{fsoto}\cite{fsoto2} 
derived from the Hubble Deep Field\cite{HDF}. 
The answer to the first question is: very little ($<1$\%), and to the second: 
a small but non-negligible amount (10\% at $b_{J} = 24$ mags rising to
30\% at $b_{J} = 28$ mags). Hence the 
cosmological significance of dwarf galaxies, from a purely baryon-centric 
perspective, lies in their contribution to the formation and assembly phase of 
the giant galaxies - where perhaps they did once dominate the baryon budget.
\end{moriondabstract}

\section{Introduction}
Dwarf galaxies have, on the whole, been overlooked in the 
grand scheme of things, often considered either insignificant or irrelevant 
for cosmological purposes. Recently though problems in the interpretation of 
ultra-deep images have brought dwarf galaxies to the fore. This conundrum,
known as the faint blue galaxy problem\cite{kk}\cite{ellis}, could be
readily explained if the local space density of dwarf galaxies was higher 
than previously supposed\cite{dpdmd} and/or dwarf galaxies underwent dramatic 
recent evolution\cite{pd}. Certainly the constraints on the space density of 
dwarf galaxies are weak. This stems from the fact that 
bright magnitude limited redshift surveys fail to probe representative volumes 
for dwarf systems\cite{dp} and that faint redshift surveys are too small, too 
incomplete, model dependent and still not faint enough\cite{dc}. An additional 
and probably more fundamental constraint is that dwarf galaxies are typically 
of low surface brightness, making spectroscopic redshift determination itself 
problematical. If their space density is high could they, (a) constitute a 
significant number of baryons (and/or cold dark matter), and (b) present 
a foreground screen of objects contaminating our window into the distant 
Universe ?

\section{The contribution of dwarf galaxies to the baryon budget}
To address this question we utilise the photometric redshift 
estimates\cite{fsoto2} for all galaxies in the Hubble Deep Field\cite{HDF} to 
$b_{J} = 28$. While photometric redshifts are less precise
than spectroscopic, they can probe to both very faint fluxes and very faint 
surface brightness, overcoming the primary spectroscopic obstacles.
The HDF photometric redshifts have been tested and verified to 
$b_{J}=26$\cite{hogg}. From this dataset we can construct a 
{\it volume limited} sample unhindered by surface brightness selection 
effects and derive an unbiased measure of the luminosity function of galaxies 
over a broad absolute magnitude range. Of course the HDF is not the ideal 
local survey instrument and to construct a ``local'' luminosity function one 
must relax the definition of ``local''. In this case to the redshift interval, 
$0.1 < z < 0.5$ resulting in a survey volume (for galaxies with $M_{B} < -13$) 
of $200$h$^{-1}$Mpc$^{3}$ for $q_{o}=0.5, \Omega=1$.
Figure 1 shows the recovered ``local'' luminosity distribution for our 
HDF sample and superimposed is the recent measurement by Loveday\cite{love}.
The Loveday fit has been scaled up by a factor of 2 which is coincidentally 
a similar factor adopted in most faint galaxy models to overcome the 
often overlooked local normalisation problem\cite{dwg}; this reopens the 
question as to whether we live in a large local 
underdensity\cite{shanks}\cite{zucca} !
The two datasets on Figure 1 (upper) 
essentially agree in shape within our volume 
limited region and then diverge; both surveys suffering from limiting 
statistics (at $M_{B} > -13$). 

\begin{figure}[h]
\centerline{\psfig{file=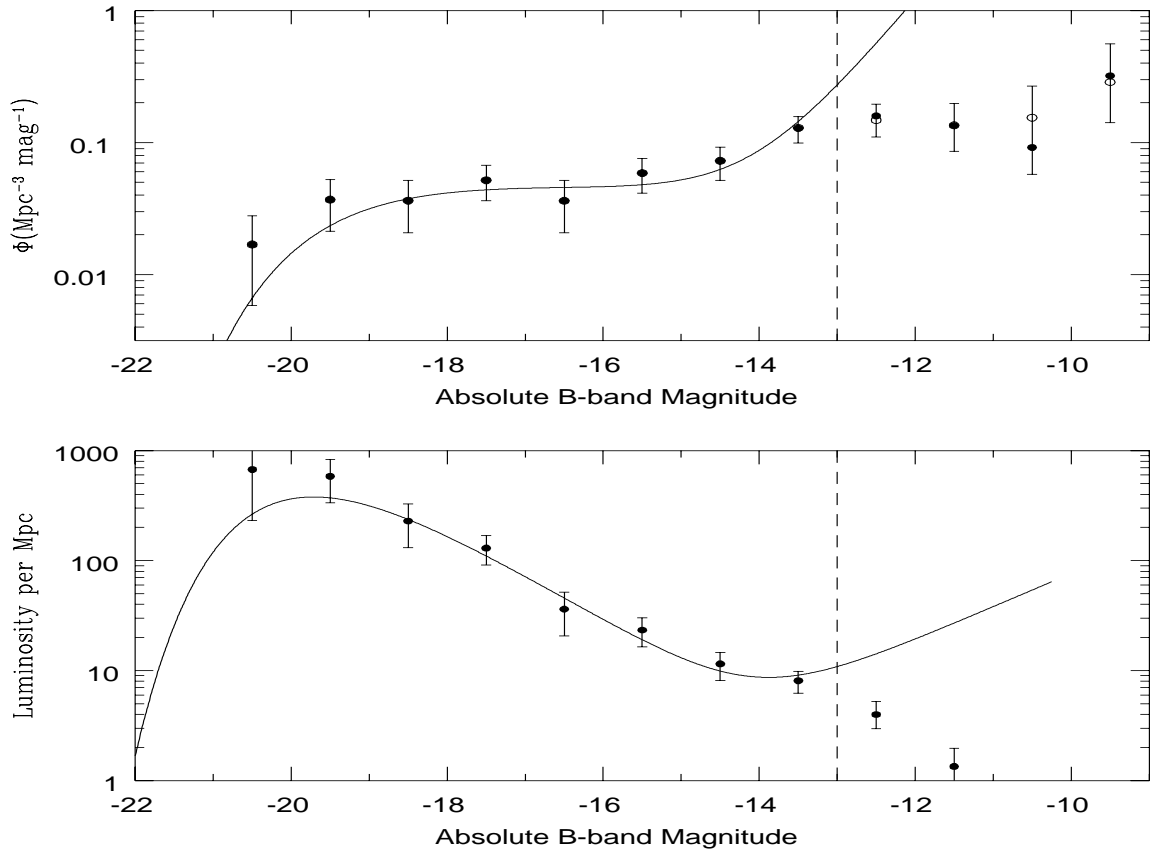,height=12.0cm,width=17.0cm}}
\caption{The local luminosity function of galaxies (upper) and the
luminosity density distribution (lower) derived from our HDF data.
The dashed line indicates the reliable volume-limited region. The solid line
shows the fit to the Loveday data (these proceedings).
If mass-to-light ratios are invariant with luminosity the lower plot
equates to the baryon density distribution.}
\end{figure}

\noindent
Figure 1 (lower) shows the total luminosity density for each 
absolute magnitude interval. To convert this to the contribution to the mass
density one requires mass-to-light ratios, an important topic for which 
relatively little data exists. Results presented at this conference do however
suggest comparable mass-to-light ratios to the giants. While this requires
substantial further work the implication is that Figure 1 (lower) translates
directly to the relative contribution to the baryon density. Dwarf galaxies 
therefore appear to contribute less than 1\% of the total field galaxy baryon 
budget for $z < 0.5$. To contribute equally to the field baryon budget from
dwarf galaxies require a mass-to-light ratio $100 \times$ higher than giants.
Two further comments: Firstly at very faint
luminosities the Loveday data is actually divergent in mass albeit in 
disagreement with our HDF constraints; secondly hierarchical merger models
would predict a shift in the peak of this distribution at higher redshifts
towards lower absolute magnitudes.

\section{The contamination of deep images by dwarf galaxies}
Figure 1 implies that while dwarf galaxies are more numerous than the 
giants this is insufficient in themselves to contribute substantially
to the faint blue galaxy excess\cite{spdhdf}. 
However if they strongly evolve they might
still play a part. Normally this is modeled by adopting a local luminosity 
function, some evolutionary scheme and a cosmological framework. However it is 
possible to quantify this directly from our dataset by calculating what 
fraction lies above or below any specified absolute magnitude. Figure 2 shows 
this result for the $B$ (left) and $I$ (right) bands respectively. 

\begin{figure}[h]
\vspace{-6.5cm}
\centerline{\psfig{file=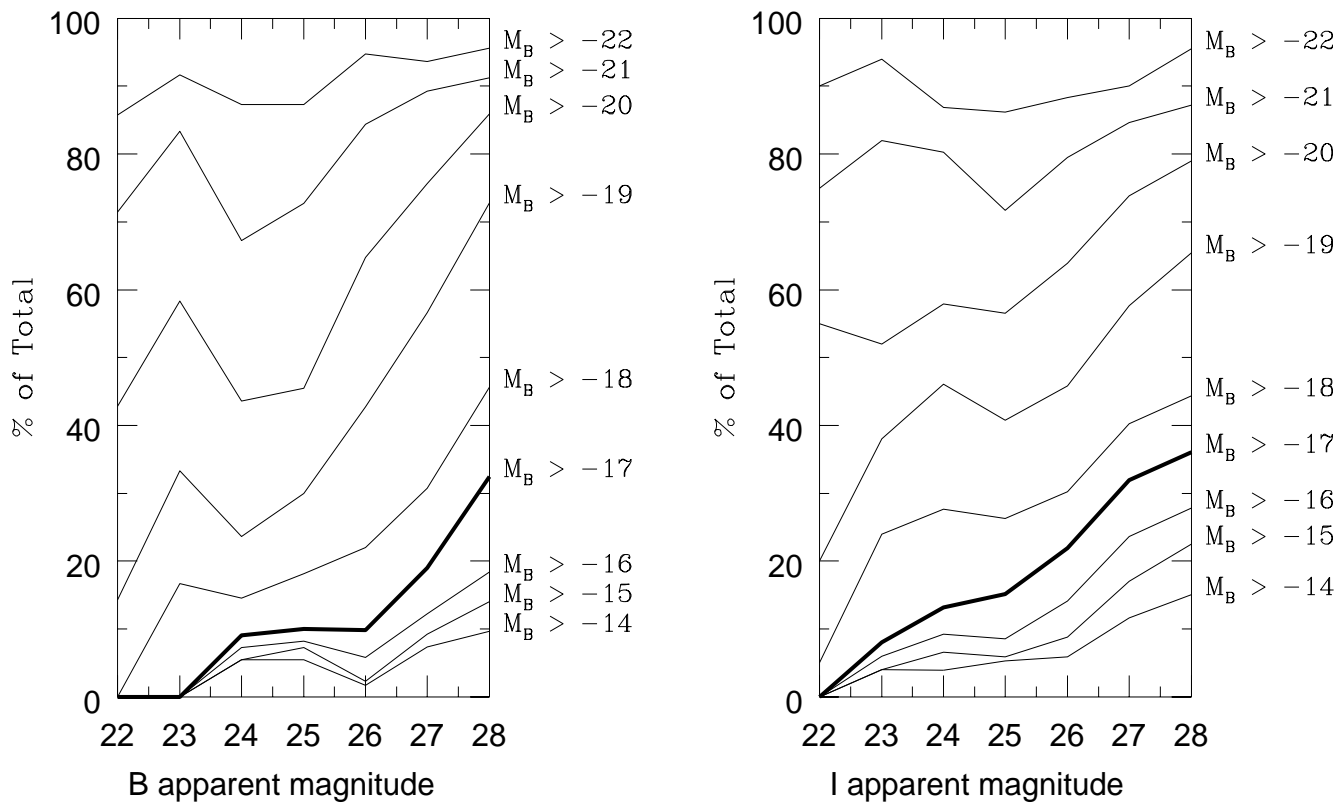,height=18.5cm,width=18.0cm}}
\vspace{-1.0cm}
\caption{The fractional contribution of galaxies below a specified absolute
magnitude to the faint galaxy counts in $B$ (right) and $I$ (left). The thick
solid line shows our preferred delineation between dwarfs and giants}
\end{figure}

\noindent
These rather unconventional plots are worth sparing a few moments
to glance over as they directly answer the initial question while circumventing
the need for any model assumptions. Shown on the y-axis is a simple percentage,
along the x-axis the apparent magnitude. The lines represent the fractional 
contribution from galaxies below a specified absolute magnitude, as indicated 
on the right hand side. One can choose one's own definition of dwarf galaxies, 
here we highlight (thick solid line) the fractional contribution from those 
galaxies with $M_{B} > -17$, our definition of dwarfs. We therefore see that 
dwarfs constitute $\sim 10$\% of the faint galaxy population by $b_{J}=24$ mags
and $I=23.5$ mags. This rises to 30\% by $b_{J}=28$ mags. The deeper we look 
the greater the contribution/contamination. Note that the distance between two 
lines on Fig. 2 indicates the contribution from a specific luminosity interval.

\section{Discussion}
It seems that dwarf galaxies represent a small fraction ($<1$\%) of the local 
($z < 0.5$) field 
galaxy baryon budget and contribute only as a minor player ($<10$\% at 
$b_{J}=24$ mags) in the faint blue galaxy problem\footnote{Of course many of 
the $M_{B} < -17$ objects may eventually fade into dwarf galaxies locally but 
if we are consistent in our definition of dwarfs galaxies they do not qualify 
at that moment in time.}. These are important results
but should not equate to the dismissal of dwarf galaxies from the domain of 
cosmology. They are still more numerous that the giants and feature critically
in hierarchical formation scenarios. In fact it seems perplexing how efficient 
is that process (gravitational instability) which transforms the smooth baryon 
distribution seen in the Cosmic Background Radiation to the lumpy local
distribution of the luminous galaxy populations. The most likely intermediary 
phase must still surely be the dwarf galaxy but their era of pre-eminence has 
long passed. We thank the staff, engineers and astronauts involved in the 
operation and maintenance of the Hubble Space Telescope and note that the 
Hubble Deep Field was obtained as a public service to the community under the 
directive of Bob Williams.

\vfill
\end{document}